\documentclass[]{mn2e}
\usepackage{graphicx}
\usepackage{epsfig}
\usepackage{amsmath}
\newcommand       \um           {\mu{\rm m}}
\newcommand       \mum        {\,{\rm \mu m}}

\newcommand       \simali       {{\sim}\,}
\newcommand       \magni        {\,{\rm mag}}
\newcommand	\gtsim	{\lower.5ex\hbox{$\buildrel > \over \sim$}}
\newcommand     \ltsim  {\lower.5ex\hbox{$\buildrel < \over \sim$}}
\newcommand	\simgt	{\lower.5ex\hbox{$\buildrel > \over \sim$}}
\newcommand     \simlt  {\lower.5ex\hbox{$\buildrel < \over \sim$}}

\newcommand       \Angstrom     {\,{\rm \AA}}
\newcommand       \g            {\,{\rm g}}
\newcommand       \cm           {\,{\rm cm}}

\newcommand       \s            {\,{\rm s}}
\newcommand       \yr           {\,{\rm yr}}
\newcommand       \erg          {\,{\rm erg}}
\newcommand       \eV           {\,{\rm eV}}
\newcommand       \nH           {n_{\rm H}}

\newcommand       \rmH          {\,{\rm H}}

\newcommand       \Cabs         {C_{\rm abs}}

\newcommand       \K            {\,{\rm K}}
\newcommand       \mH         {m_{\rm H}}

\newcommand{\RV}{R_{\rm V}}

\newcommand       \cgas     {\left[{\rm C/H}\right]_{\rm gas}}

\newcommand       \sidust   {\left[{\rm Si/H}\right]_{\rm dust}}
\newcommand       \cdust    {\left[{\rm C/H}\right]_{\rm dust}}

\newcommand       \mgdust   {\left[{\rm Mg/H}\right]_{\rm dust}}

\newcommand       \Cs      {C_{\rm s}}
\newcommand       \Cg      {C_{\rm g}}
\newcommand       \alphas  {\alpha_{\rm s}}
\newcommand       \alphag  {\alpha_{\rm g}}
\newcommand       \betas   {\beta_{\rm s}}
\newcommand       \betag   {\beta_{\rm g}}
\newcommand       \ats     {a_{\rm t,s}}
\newcommand       \atg     {a_{\rm t,g}}
\newcommand       \acs     {a_{\rm c,s}}
\newcommand       \acg     {a_{\rm c,g}}

\newcommand       \ppm     {\,{\rm ppm}}

\newcommand       \bICE    {b_{\scriptscriptstyle\rm ICE}}
\newcommand \sisun {\left[{\rm Si/H}\right]_\odot}
\newcommand \csun  {\left[{\rm C/H}\right]_\odot}

\newcommand \sistar {\left[{\rm Si/H}\right]_\star}
\newcommand \cstar  {\left[{\rm C/H}\right]_\star}

\newcommand \water  {{\rm H_2O}}
\newcommand \taupd  {\tau_{\rm pd}}
\newcommand \taumc  {\tau_{\rm MC}}
\newcommand \Ypd    {Y_{\rm pd}}
\newcommand \muice  {\mu_{\scriptscriptstyle\rm ICE}}
\newcommand \rhoice {\rho_{\scriptscriptstyle\rm ICE}}
\newcommand \Ndot   {\dot{N}}
\newcommand \sgrA {{\rm Sgr}\,{\rm A}^{\ast}}
\newcommand       \mnras        {MNRAS}
\newcommand       \apj          {ApJ}
\newcommand       \apjs         {ApJS}
\newcommand       \apjl         {ApJL}

\newcommand       \aap        {A\&A}
\newcommand       \araa         {ARA\&A}
\title{
   The Interstellar Oxygen Crisis, or Where Have
   All the Oxygen Atoms Gone?$^{}$\thanks{Dedicated to 
   the late Professor J.~Mayo~Greenberg (1922.1.14--2001.11.29)
   of Leiden University who suggested the possible 
   existence of interstellar snowballs four decades ago.} 
   }

\author[Wang, Li, \& Jiang]
       {Shu Wang$^{1,2}$\thanks{shuwang@mail.bnu.edu.cn},
        Aigen Li$^{2}$\thanks{lia@missouri.edu}, and
        B.W.~Jiang$^{1}$\thanks{bjiang@bnu.edu.cn}\\
        $^1$Department of Astronomy,
            Beijing Normal University,
            Beijing 100875, China\\
        $^2$Department of Physics and Astronomy,
             University of Missouri,
             Columbia, MO 65211, USA
             }
\begin{document}
\date{Received data: June 29 2015/ Accepted date:  August 13 2015}
\pagerange{\pageref{firstpage}--\pageref{lastpage}} \pubyear{2015}

\maketitle

\label{firstpage}
\begin{abstract}
The interstellar medium (ISM) seems to have a significant
{\it surplus} of oxygen 
which was dubbed as the ``O crisis'':
independent of the adopted interstellar reference abundance,
the total number of O atoms 
depleted from the gas phase
far exceeds that tied up in solids
by as much as $\simali$160$\ppm$ of O/H.
Recently, it has been hypothesized that the missing O
could be hidden in micrometer-sized H$_2$O ice grains.
We examine this hypothesis
by comparing the infrared (IR) extinction
and far-IR emission arising from
these grains
with that observed in the Galactic diffuse ISM.
We find that it is possible for the diffuse ISM
to accommodate $\simali$160$\ppm$ of O/H
in $\mu$m-sized $\water$ ice grains
without violating the observational constraints
including the absence of the 3.1$\mum$ O--H absorption feature.
More specifically, $\water$ ice grains of
radii $\simali$4$\mum$ and O/H\,=\,160$\ppm$
are capable of accounting for
the observed flat extinction
at $\simali$3--8$\mum$
and produce no excessive emission
in the far-IR.
These grains could be present in the diffuse ISM
through rapid exchange of material
between dense molecular clouds
where they form
and diffuse clouds
where they are destroyed by
photosputtering.
\end{abstract}



\section{Introduction\label{sec:intro}}
The depletion of certain heavy elements
from the interstellar gas,
known as the ``interstellar depletion'',
was first noticed by Morton et al.\ (1973)
who found that the gas-phase abundances of
these elements measured by
the {\it Copernicus} ultraviolet (UV) satellite
for interstellar clouds are significantly
lower than in the Sun.
Based on the strong correlation
between the depletions of these elements
and the condensation temperatures
at which they are incorporated into
solid grains in stellar atmospheres and nebulae,
Field (1974) proposed that the elements missing
from the gas phase must have condensed into dust grains.

Assuming the interstellar reference abundances
to be the same as the accepted solar abundances
at the time
(Cameron 1973),
Greenberg (1974) compared
the observed interstellar depletions
with the abundances of the dust-forming elements
required to account for the observed visual extinction.
He found that a substantial amount of
the intermediate-weight elements
O, C, and N was unaccounted for
by the interstellar gas and dust
whereas the heavier elements
Si, Mg and Fe were underabundant.

Snow \& Witt (1996) compiled the elemental abundances
of early B stars and young F and G stars
and found that their abundances are just
$\simali$50--70\% of the then accepted solar abundances
of Anders \& Grevesse (1989).
They argued that, due to their young ages,
the photospheric abundances of unevolved B stars
and young F and G stars
are more representative of
the interstellar composition
than the 4.6-billion-year-old Sun.
If the interstellar abundances are indeed subsolar,
the original question of a surplus of
missing, unaccountable elements
posed by Greenberg (1974) then turned into
a shortage of raw materials for making the dust
to account for the observed extinction.
This was most pronounced for carbon.
Whereas published values for the Sun
range from $\csun\approx350\ppm$
to $\approx$\,470$\ppm$,
where ppm stands for parts per million,
Snow \& Witt (1995) derived a C/H abundance
of $\simali$225$\ppm$ for Galactic stars.
With the gas-phase abundance of
$\cgas\approx140\ppm$ (Cardelli et al.\ 1996) or
$\approx$\,100$\ppm$ (Sofia et al.\ 2011)\footnote{%
  Jenkins (2014) argued that the gas-phase C/H abundance
  of Sofia et al.\ (2011) derived from
  the strong transition of C\,II at 1334$\Angstrom$ 
  may be more trustworthy than
  the previous values measured from
  the weak intersystem absorption
  transition of C\,II] at 2325$\Angstrom$.
  }
subtracted, the remaining C/H abundance of
$\simali$85$\ppm$ or $\simali$125$\ppm$
is insufficient to form the carbonaceous dust
required by all dust models.
This, known as the ``C crisis'', still holds
when one compares the most recent determinations
of the solar abundance
of C/H\,$\approx$\,269$\ppm$ 
(Asplund et al.\ 2009)
and early B stars
of C/H\,$\approx$\,214$\ppm$ 
(Nieva \& Przybilla 2012)\footnote{%
   Poteet et al.\ (2015) found that the B-star
   Si and Mg abundances are not enough to account 
   for the 9.7$\mum$ Si--O absorption feature
   observed toward $\zeta$ Ophiuchi
   (see their Figure~8).
   }
with the latest dust models
(e.g., Jones et al.\ 2013)
which all require $\cdust>200\ppm$
to be locked up in carbonaceous dust.
Note that the refined studies of the solar abundances
since Anders \& Grevesse (1989) all led to a gradual,
downward revision such that the derived solar abundances
were no longer substantially higher than that of B stars,
and hence
the C crisis remains unalleviated
with the new solar and B-star abundances.
    It is worth noting that Lodders (2003) argued that 
    the currently observed solar photospheric abundances 
    must be lower than those of the proto-Sun 
    because heavy elements have settled toward the Sun's 
    interior since the time of the Sun's formation 
    some 4.6\,Gyr ago.
    She further argued that the protosolar abundances 
    are more representative of 
    the solar system elemental abundances.

The depletion of oxygen is also problematic.
Unlike C of which the depletion is insufficient
to account for the observed extinction,
the major solid-phase reservoirs of O
in the diffuse interstellar medium (ISM)
--- silicates and metal oxides ---
are insufficient to account for
the total O/H abundance missing from
the gas phase. Jenkins (2009)
examined the relative proportions of
17 individual elements that are incorporated into dust
along 243 different Galactic sightlines.
He found that the depletion of O
in the diffuse ISM far exceeds
the consumption of O by silicates and metallic oxides,
except for low-density regions with little depletions.
This is insensitive to the adopted reference
abundance since the approach taken by Jenkins (2009)
was based on differential depletions rather than
absolute values.
Whittet (2010b) investigated the depletions of O
over a wide range of environments
from the tenuous intercloud medium
and diffuse clouds to dense clouds
where H$_2$O ice is present.
He found that as much as
$\simali$160$\ppm$ of O/H
is unaccounted for at the interface
between diffuse and dense phases, again,
independent of the choice of reference abundances.
This surplus of O,
dubbed as the interstellar ``O crisis''
by Whittet (2010a),
poses a severe challenge
to our understanding of interstellar dust.

The nondetection of the 3.1$\mum$ O--H stretching
absorption feature of H$_2$O ice in the diffuse ISM
rules out submicrometer-sized H$_2$O ice grains
as a significant reservoir of O
(Whittet et al.\ 1997, Poteet et al.\ 2015).
Jenkins (2009) suggested that large amounts of O
could be hidden in H$_2$O ice grains
(or large grains with thick mantles of H$_2$O ice)
that have diameters of the order of or greater than $1\mum$.
Grains this large will have the 3.1$\mum$ O--H feature
substantially suppressed.\footnote{%
  As illustrated in Figure~4 of Poteet et al.\ (2015),
  for 2.0--3.2$\mum$-sized H$_2$O ice grains,
  with increasing grain size,
  both the scattering and absorption 
  at the 3.1$\mum$ O--H stretch decrease.
  }
More recently, Poteet et al.\ (2015) analyzed
the $\simali$2.4--36$\mum$ absorption spectrum
of the sightline toward $\zeta$ Ophiuchi
obtained with
the {\it Short Wavelength Spectrometer} (SWS)
on board the {\it Infrared Space Observatory} (ISO)
and the {\it Infrared Spectrograph} (IRS)
on board the {\it Spitzer Space Telescope}.
They determined the elemental abundances of 
O, Mg, and Si in silicates.
Along with the upper limit estimates 
of O in other materials, they found that
as much as $\simali$156$\ppm$ of O/H 
is unaccounted for along the line of sight
toward $\zeta$ Ophiuchi,
a prototypical cool diffuse cloud
and an environment
near the diffuse-dense ISM transition.
They argued that the missing reservoir of O
must reside on very large, micrometer-sized grains
(e.g., with radii $a\simgt3.2\mum$) 
which are nearly opaque to infrared (IR) radiation.

The aim of this work is to test the hypothesis
of $\mu$m-sized H$_2$O ice grains as the repository
of the unaccounted for $\simali$160$\ppm$ of O/H in
the diffuse ISM.
Roughly speaking, dust scatters and absorbs starlight
most effectively at a wavelength $\lambda$ comparable
to its spherical radius $a$: $4 a (n-1)/\lambda\sim1$,
where $n(\lambda)$ is the real part
of the index of refraction of the dust.
  This can be understood in terms of elementary optics:
  the maxima of the extinction could be
  considered as being caused by the destructive interference
  between the incident and forward-scattered light.
  The phase difference between a ray that traverses
  a large transparent sphere without deviation
  (i.e., the forward-scattered ray)
  and a ray that traverses the same physical path
  outside the sphere is
  $\Delta\phi = \left(2\pi/\lambda\right) 2 a (n-1)$.
  The maxima of the extinction
  occur at $\Delta\phi = (2j+1)\pi$,
  where $j$ is an integer,
  with $j=0$ corresponding to
  the principal extinction peak
  (see Bohren \& Huffman 1983).

For H$_2$O ice, $n\approx1.3$ at $\lambda\gtsim1\mum$
(Warren 1984), thus, $\mu$m-sized H$_2$O ice grains
will cause extinction at $\lambda>1\mum$.
Therefore, we will examine in \S\ref{sec:irext}
the consistency between the observed IR extinction
and the extinction resulting from the standard
silicate-graphite model with the addition
of a population of $\mu$m-sized H$_2$O ice grains
of O/H\,=\,160$\ppm$.
On the other hand, H$_2$O ice grains are transparent
in the ultraviolet (UV) and visible wavelength range
and hence they will be cold and emit in the far-IR
at $\lambda\gtsim200\mum$.
Therefore, we will examine in \S\ref{sec:irem}
the IR radiation emitted by $\mu$m-sized H$_2$O
ice grains and make a comparison between the model
IR emission and that observed for the Galactic
diffuse ISM.
Their origin and survival will be discussed
in \S\ref{sec:discussion}.
The major conclusions are summarized
in \S\ref{sec:summary}.

\begin{figure*}
\vspace{-6mm}
\centering
\includegraphics[angle=0,width=20cm]{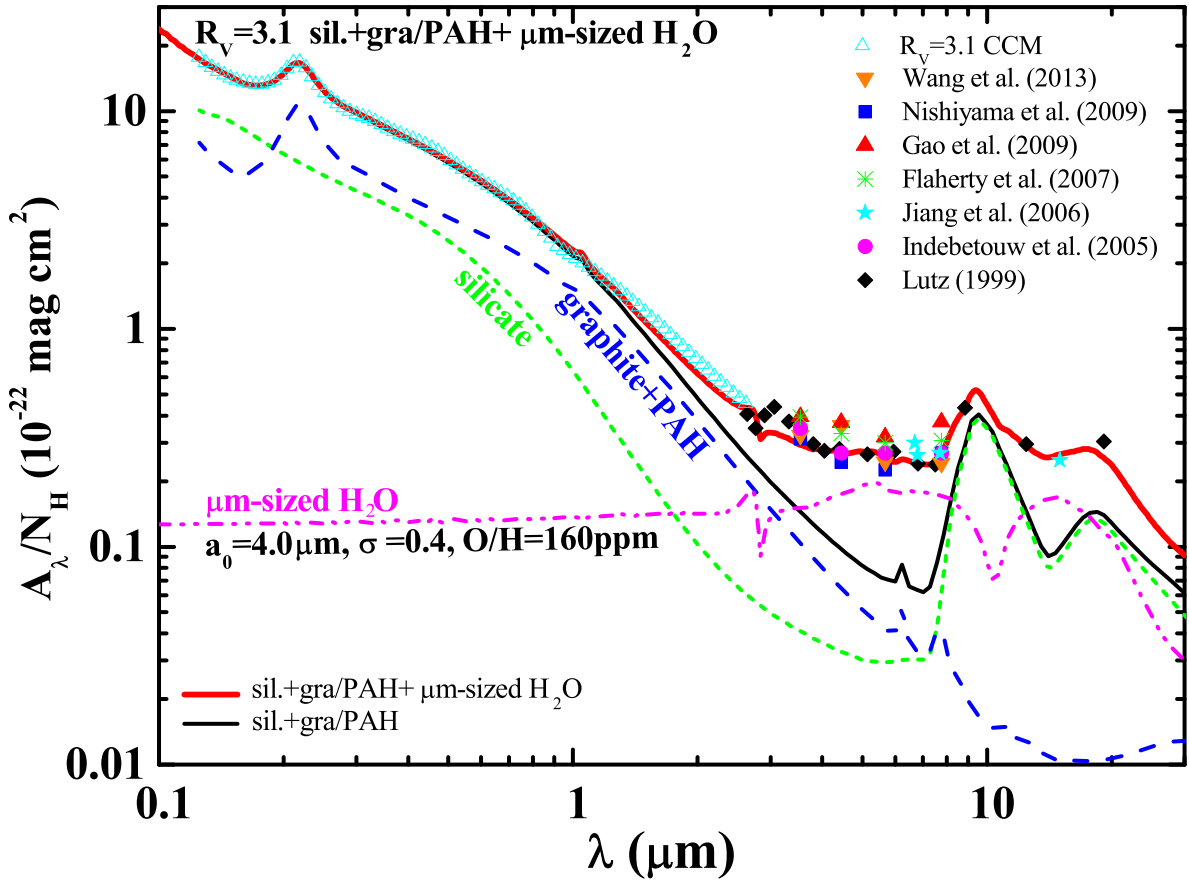}
\vspace{-11mm}
\caption{
          \label{fig:irext}
          Fitting the $\RV=3.1$ extinction curve
          from the UV/optical to the near- and mid-IR
          with (1) amorphous silicate
          (green short dashed line),
          (2) graphite and PAHs (blue dashed line), and
          (3) $\mu$m-sized H$_2$O ice grains
          (magenta dash-dot-dotted line)
          of O/H\,=\,$160\ppm$
          with a log-normal size distribution
          characterized by $a_0\approx4.0\mum$ and
          $\sigma\approx0.4$.
          The thick red solid line plots the model fit
          which is the sum of silicate, graphite/PAHs,
          and $\mu$m-sized H$_2$O ice.
          The black solid line plots the sum of silicate
          and graphite/PAHs.
          The symbols plot the observed extinction:
          the cyan open triangles plot the $\RV=3.1$
          UV/optical/near-IR extinction,
          and the other symbols plot
          the mid-IR extinction (see text).
          }
\vspace{-5mm}
\end{figure*}

%
\section{Extinction}\label{sec:irext}
For $\water$ ice grains to be considered as
a viable component of the diffuse ISM,
they should satisfy the constraints
placed by the observed extinction
from the far-UV to the IR,
including the absence of the 3.1$\mum$
absorption feature of $\water$ ice.
The observed UV/optical interstellar extinction
can be characterized by
a single parameter $R_V$
(Cardelli et al.\ 1989, CCM).\footnote{%
  $R_V\,\equiv\,A_V/\left(A_B-A_V\right)$
  is the total-to-selective extinction ratio,
  where $A_B$ and $A_V$ are
  the blue- and visual-band extinction
  at $\lambda_B=4400\Angstrom$
  and $\lambda_V=5500\Angstrom$, respectively.
  For the Galactic average, $R_V\approx3.1$.
  }
As elaborated in Wang, Li \& Jiang (2014),
while the UV/optical extinction can be closely
fitted by the classical silicate-graphite model
(Mathis et al.\ 1977, Draine \& Lee 1984),
this model predicts a power-law
of $A_\lambda \propto \lambda^{-1.75}$
at $1\mum < \lambda < 7\mum$ (Draine 1989)
which is too steep to be consistent
with the subsequent observations made by
{\it ISO} and {\it Spitzer}
since numerous observations suggest that
the mid-IR extinction at $3\mum <\lambda< 8\mum$
is flat or ``gray'' for both diffuse
and dense environments (see Figure~\ref{fig:irext}),
including the Galactic center
(Lutz 1999, Nishiyama et al.\ 2009), 
the Galactic plane
(Indebetouw et al.\ 2005,
Jiang et al.\ 2006, Gao et al.\ 2009),
the Coalsack nebula (Wang et al. 2013),
and nearby star-forming regions (Flaherty et al.\ 2007).
All these observations appear to suggest
an ``universally'' flat mid-IR extinction law,
with little dependence on environments.

We construct an ``observed'' extinction curve
for the diffuse ISM as follows:
(i) for $0.125\mum < \lambda < 3\mum$,
we take the Galactic average of
$R_V=3.1$ as parameterized by CCM
and divide it into 95 wavelengths,
equally-spaced in $\ln \lambda$;
(ii) for the mid-IR extinction
at $3\mum < \lambda < 8\mum$,
we first obtain a weighted ``average''
from the observed extinction
shown in Figure~\ref{fig:irext},
with twice as much weight given to
the diffuse sightlines toward
the Galactic center,\footnote{%
   The extinction along the line of sight
   toward the Galactic center is believed 
   to be dominated by dust in the diffuse ISM
   as revealed by the detection of the 3.4$\mum$
   aliphatic C--H absorption feature
   which is absent in dense regions
   (Sandford et al.\ 1991, Pendleton et al.\ 1994, 
    Tielens et al.\ 1996).
   However, it does contain molecular cloud materials 
   as revealed by the detection of the 3.1 and 6.0$\mum$ 
   $\water$ ice absorption features, 
   e.g., the sightline toward the Galactic center source
   $\sgrA$ suffers $\simali$30$\magni$ of
   visual extinction (McFadzean et al.\ 1989),
   to which molecular clouds may contribute as much as
   $\simali$10$\magni$ (Whittet et al.\ 1997).
   As shown in Figure~\ref{fig:irext}, 
   the $\simali$3--8$\mum$ extinction curve 
   of the Galactic center 
   (Lutz 1999, Nishiyama et al.\ 2009)
   is less flatter than that of other regions.
   When constructing the ``observed'' extinction curve,
   if one assigns less weight to the Galactic center
   sightlines, one would obtain an even flatter
   $\simali$3--8$\mum$ extinction law.
   Therefore, even more O/H could be tied up
   in $\mu$m-sized $\water$ ice grains.
   }
we then interpolate the ``average'' mid-IR extinction
into 25 logarithmically equally-spaced wavelengths.

We aim at reproducing the observed extinction
from the UV/optical to the near- and mid-IR
with a mixture of amorphous silicate dust
and carbonaceous dust as well as $\water$ ice grains.
The UV/optical extinction is predominantly caused
by sub-$\mu$m- and nano-sized grains (Li 2004).
The nondetection of the 3.1$\mum$ absorption feature
of $\water$ ice in the diffuse ISM
places an upper limit of 
$\simali$$2\ppm$ (Whittet et al.\ 1997)
and $\simali$$9\ppm$ (Poteet et al.\ 2015)
of O/H in sub-$\mu$m-sized $\water$ ice grains
toward Cyg OB2 No.\,12 and $\zeta$ Oph, 
respectively.
Therefore, the observed UV/optical extinction
is mainly produced by the silicate and carbonaceous
components while the $\water$ ice component,
if present in the diffuse ISM, must be larger
than $\simali$1$\mum$ and cause extinction
at $\lambda\gtsim1\mum$.
%

Following Weingartner \& Draine (2001; WD01),
we model the observed extinction
in the wavelength range of
$0.125\mum < \lambda < 8\mum$
in terms of the silicate-graphite-PAH model
combined with a population of $\mu$m-sized
$\water$ ice grains.
We take the size distribution functional form
of WD01 for the silicate and graphitic component,
assuming the latter extends
from grains with graphitic properties
at radii $a\gtsim50\Angstrom$,
down to grains with PAH-like properties
at very small sizes (see Li \& Draine 2001).
The dielectric functions of amorphous silicate,
graphite, and $\water$ ice are taken from
Draine \& Lee (1984) and Warren (1984).
The WD01 model employs
two log-normal size distributions
for two populations of PAHs
which respectively peak at $a_{0,1}$, $a_{0,2}$
and have a width of $\sigma_{1}$, $\sigma_{2}$,
consuming a C abundance of
$b_{{\rm C},1}$, $b_{{\rm C},2}$ (per H nuclei).
Following Draine \& Li (2007),
we adopt $a_{0,1}=3.5\Angstrom$, $\sigma_1=0.40$,
$b_{\rm C,1}=45\ppm$, $a_{0,2}=20\Angstrom$,
$\sigma_2=0.55$, and $b_{\rm C,2}=15\ppm$.
These parameters were constrained
by the observed near- and mid-IR emission.
For the $\mu$m-sized $\water$ ice component,
we also adopt
a log-normal size distribution
of peak size $a_0$ and width $\sigma$:
\begin{equation}\label{eq:icednda}
\begin{aligned}
\frac{1}{\nH} \frac{dn}{da}=
&
\frac{3}{(2\pi)^{3/2}}
\times
\frac{\exp\left(-4.5\sigma^2\right)}{\rhoice a_0^3 \sigma}
\times
\frac{\bICE\muice\mH}{2} \\
&
\times
\frac{1}{a}\exp\left\{-\frac{1}{2}
\left[\frac{\ln(a/a_0)}{\sigma}\right]^2\right\}, \\ ~~
\end{aligned}
\end{equation}
where $\mH$ is the atomic H mass,
$\rhoice$ and $\muice$ are respectively the mass density
and the molecular weight of $\water$ ice
($\rhoice\approx1.0\g\cm^{-3}$ and $\muice\approx18$),
and $\bICE$ is the O abundance per H nuclei locked up
in $\mu$m-sized $\water$ ice grains.

Following WD01,
we use the Levenberg-Marquardt method
(Press et al.\ 1992) 
to minimize the fitting error
between the observed and model extinction
(see Wang, Li \& Jiang 2015).
As shown in Figure~\ref{fig:irext},
together with silicate, graphite and PAHs,
$\water$ ice grains with
$a_0\approx\left(4.0\pm1.0\right)\mum$,
$\sigma\approx\left(0.4\pm0.1\right)$,
and $\bICE=160\ppm$ of O/H
satisfactorily reproduce the $\simali$3--8$\mum$
mid-IR extinction.
Figure~\ref{fig:irext} also shows that
these $\mu$m-sized $\water$ ice grains
are ``gray'' in the UV/optical and make
negligible contribution to the observed
UV/optical extinction.
The size distributions of
the silicate and graphite components
which closely reproduce the observed
UV/optical extinction
are characterized by
the following parameters:
$\Cg\approx5.75\times10^{-12}$,
$\alphag\approx-1.40$,
$\betag\approx0.0291$,
$\atg\approx0.00818\mum$,
and $\acg\approx0.173\mum$ for graphite,
$\Cs\approx7.56\times10^{-14}$,
$\alphas\approx-2.19$,
$\betas\approx-0.586$,
$\ats\approx0.204\mum$,
and $\acs=0.1\mum$ for silicate
(see WD01 for the definition of each parameter).

The model extinction does not show
the 3.1$\mum$ absorption feature
which is absent in the diffuse ISM.
However, it exhibits a narrow, minor
structure at $\simali$2.8$\mum$
arising from the scattering of
the O--H stretch of $\water$ ice.
The extinction toward the Galactic center
derived by Lutz (1999) based on
the recombination lines of atomic H
was not sufficiently resolved in wavelength
to rule out this structure.
We note that Whittet et al.\ (1997) reported
the detection of a shallow feature
centered at $\simali$2.75$\mum$
in the near-IR spectrum
of Cygnus OB2 No.\,12 obtained by
the {\it Kuiper Airborne Observatory} (KAO)
and {\it ISO}/SWS.
They tentatively attributed this feature
to the OH groups of hydrated silicates.
However, it was later found to be an artifact
caused by the calibration uncertainty
(see Whittet et al.\ 2001).

\begin{figure*}
\vspace{-5mm}
\centering
\includegraphics[angle=0,width=18cm]{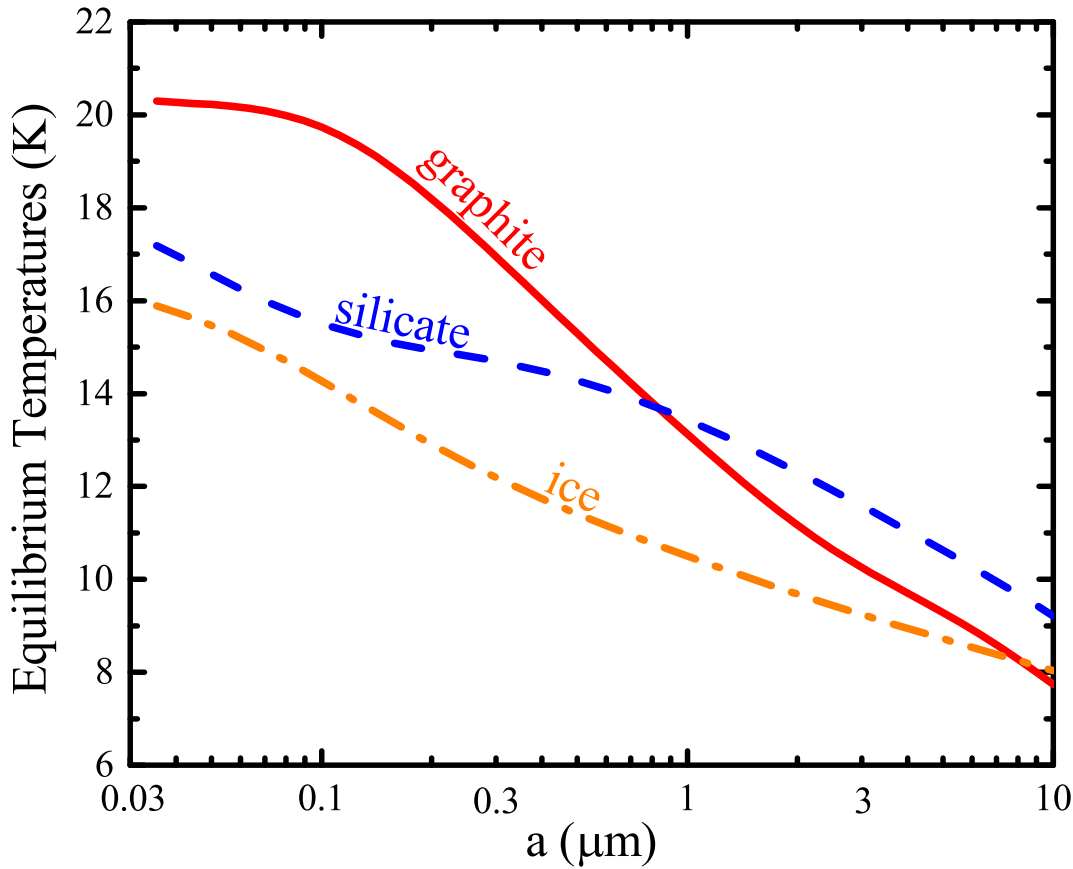}
\vspace{-12mm}
\caption{
        \label{fig:Teq}
	    Equilibrium temperatures
        for graphite (red solid line),
        silicate (blue dashed line),
        and $\water$ ice grains
        (orange dash-dotted line)
        heated by the MMP83 ISRF.
        }
\vspace{-5mm}
\end{figure*}

\begin{figure*}
\vspace{-5mm}
\centering
\includegraphics[angle=0,width=20cm]{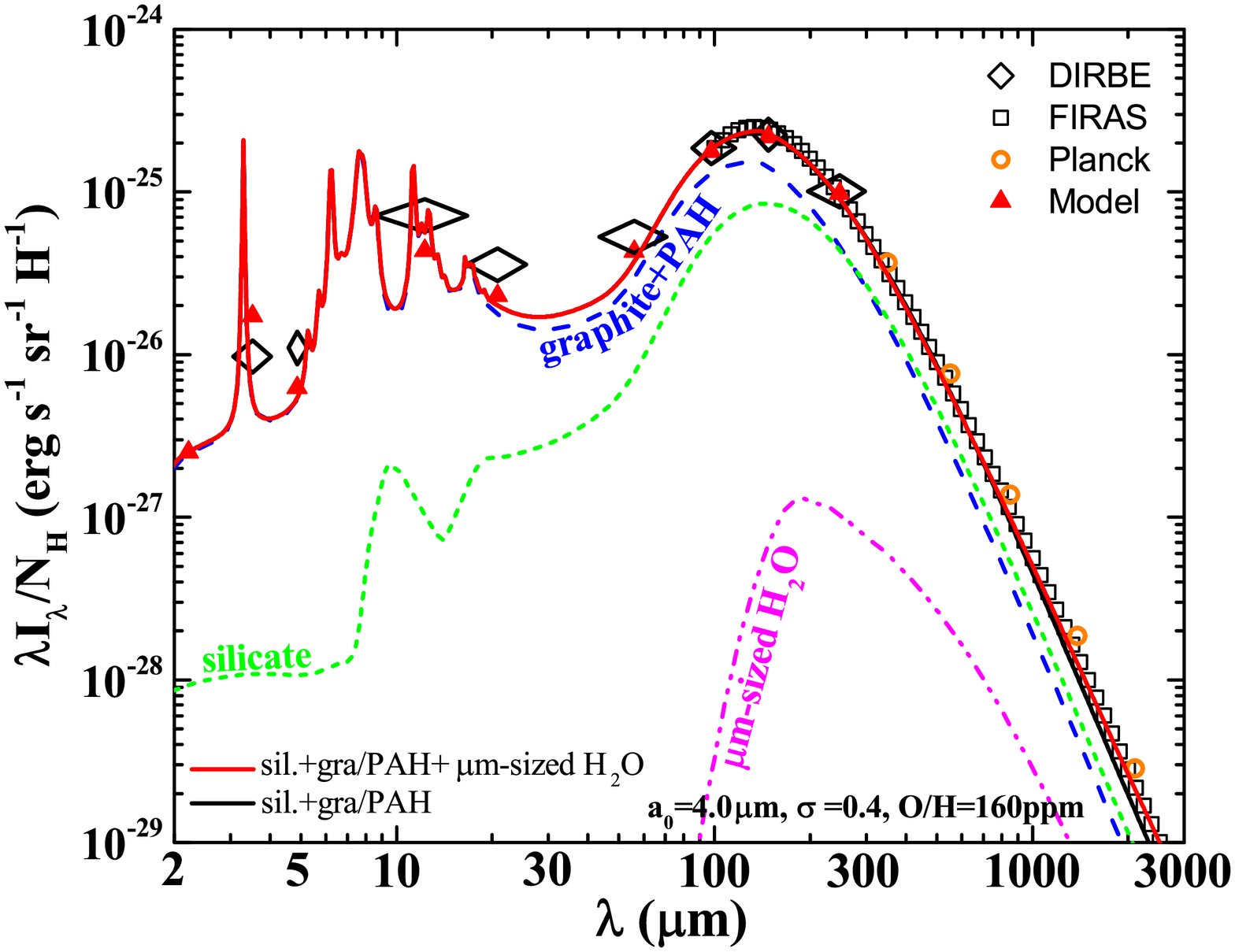}
\vspace{-14mm}
\caption{
         \label{fig:irem}
          Comparison of the observed emission
          from the diffuse ISM
          (black diamonds, black squares,
           and orange circles)
          to the model which is the sum of
          silicate (green short dashed line),
          graphite/PAHs (blue dashed line), and
          $\mu$m-sized H$_2$O ice grains
          (magenta dash-dot-dotted line).
          Red triangles show the model spectrum
          (red solid line)
          convolved with the {\it DIRBE} filters.
          Observational data are from {\it DIRBE}
          (black diamonds; Arendt et al.\ 1998),
          {\it FIRAS} (black squares; Finkbeiner et al.\ 1999),
          and {\it Planck} (orange circles;
          Planck Collaboration XVII 2014).
         }
\vspace{-5mm}
\end{figure*}

%
\section{Infrared Emission}\label{sec:irem}
To examine whether the model, with the inclusion of
a population of $\mu$m-sized $\water$ ice grains,
will result in too much emission in the far-IR,
we calculate the equilibrium temperatures
of silicate and graphite grains of radii
larger than $\simali$250$\Angstrom$
as well as $\mu$m-sized $\water$ ice grains
heated by the Mathis, Mezger, \& Panagia (1983, MMP83)
interstellar radiation field
(ISRF; see Figure~\ref{fig:Teq}).
We also calculate the temperature probability
distribution functions
of PAHs, small graphite and silicate grains
of radii smaller than $\simali$250$\Angstrom$
since the heat contents of these ultrasmall grains
are smaller than or comparable to the energy of
a single UV photon and therefore, they will be
transiently heated
and will not attain an equilibrium temperature
(see Draine \& Li 2001).

For a given size, $\water$ ice grains
are colder than silicate and graphite grains
(e.g., $T$\,$\approx$\,14.3, 15.5, 19.7$\K$
for $\water$ ice, silicate and graphite grains
of $a=0.1\mum$, respectively).
For a given composition, the dust
temperature decreases
as the grain size increases
(see Figure~\ref{fig:Teq}).
The best-fit $\water$ ice grains of radii
of $\simali$4$\mum$ have an equilibrium temperature
of $\simali$8.9$\K$.
As shown in Figure~\ref{fig:irem}, these grains
mainly emit at $\lambda\gtsim200\mum$.

Figure~\ref{fig:irem} presents the model IR emission
which combines the contributions from silicate grains,
graphite/PAH grains, and $\water$ ice grains.
We compare the model IR emission with
the diffuse ISM observed by {\it Planck}
and the {\it Diffuse Infrared Background Experiment} (DIRBE)
and the {\it Far Infrared Absolute Spectrophotometer} (FIRAS)
instruments aboard the {\it Cosmic Background Explorer} (COBE).
Figure~\ref{fig:irem} shows that the model closely fits
the observed emission
from the near-IR to the far-IR and millimeter (mm).
The $\mu$m-sized $\water$ ice component
is ``gray'' in the UV/optical
(see Figure~\ref{fig:irext})
and does not absorb much and therefore
by implication, it does not emit much
in the far-IR.
  Although the mass of the $\mu$m-sized
  $\water$ ice component exceeds that of
  graphite/PAHs by $\simali$7\%,
  the UV/optical extinction contributed
  by the former is smaller than that of
  the latter by a factor of $\simali$50.
  Moreover, the UV/optical extinction of
  $\water$ ice grains is dominated by
  scattering, instead of absorption.
The silicate-graphite/PAH-ice model
results in a total IR intensity of
$\simali$$4.48\times10^{-24}\erg\s^{-1}\rmH^{-1}$.
The fractional contributions of
silicate, graphite/PAHs, and H$_2$O ice
are approximately 27.9\%, 71.7\%, and 0.34\%,
respectively.

%
\section{Discussion}\label{sec:discussion}
We have shown in \S\ref{sec:irext}
and \S\ref{sec:irem} that the diffuse ISM
has no problem in accommodating the unaccounted for
$\simali$160$\ppm$ of O/H.
With the missing $\simali$160$\ppm$ of O/H
tied up in $\mu$m-sized $\water$ ice grains,
the silicate-graphite/PAH-ice model closely
reproduces the observed extinction
from the far-UV to the mid-IR
(see Figure~\ref{fig:irext})
as well as
the observed thermal emission
from the near-IR to
the submm/mm wavelengths
(see Figure~\ref{fig:irem}).
However, the ``C crisis'' still persists.
The model requires $\cdust\approx225\ppm$ to
be depleted in graphite/PAHs
(with $\simali$60$\ppm$ of C/H in PAHs,
see Li \& Draine 2001).
With the gas-phase C/H abundance of
$\cgas\approx140\ppm$ (Cardelli et al.\ 1996)
or $\cgas\approx100\ppm$ (Sofia et al.\ 2011)
included, the total required C/H abundance
becomes ${\rm C/H}\approx365\ppm$
or ${\rm C/H}\approx325\ppm$,
exceeding that of
solar
($\csun\approx269\pm31\ppm$, Asplund et al.\ 2009),
proto-Sun
($\csun\approx288\pm27\ppm$, Lodders 2003),
and early B stars
($\cstar\approx214\pm20\ppm$, Nieva \& Przybilla 2012).
We note that the shortage of C/H is true for
all dust models
(e.g., Zubko et al.\ [2004; ZDA] required
$\cdust\approx244\ppm$,
Jones et al.\ [2013] required
$\cdust\approx233\ppm$).
Mathis (1996) argued that porous dust
is more effective in absorbing starlight
than compact dust
and therefore one might require fewer C atoms
to account for the observed extinction.
However, Li (2005a) showed that,
through an analysis based on
the Kramers-Kronig relation of Purcell (1969),
models invoking porous dust cannot appreciably
lower the C/H consumption.
A likely solution to the C crisis problem
is that the stellar photospheric abundances
may be considerably lower than that of
the interstellar material
from which young stars are formed
(see Li 2005a).
Alternatively, if the interstellar C/H abundance
is like the {\it current} proto-solar abundance of
C/H$\simali$339$\ppm$ (Asplund 2009) 
when taking into account 
the Galactic chemical enrichment
over the past 4.6\,Gyr 
(Chiappini, Romano, \& Matteucci 2003),
with $\cgas\approx100\ppm$ (Sofia et al.\ 2011) subtracted,
the available C/H abundance ($\simali$239$\ppm$)
is approximately consistent with the latest dust models
and the so-called C-crisis will be alleviated.

The silicate-graphite/PAH-ice model requires
$\sidust\approx\mgdust\approx40.4\ppm$
to be depleted in silicate dust.
Here we assume a stoichiometric composition of
MgFeSiO$_4$ for the silicate dust which 
depletes all the Si, Mg, and Fe atoms.
For Si, it is higher than that of solar
($\sisun\approx32.4\pm2.2\ppm$,
Asplund et al.\ 2009)
and early B stars
($\sistar\approx31.6\pm3.6\ppm$,
Nieva \& Przybilla 2012),
it is consistent with
the proto-Sun abundance of
${\rm Si/H}\approx40.7\pm1.9\ppm$ 
(Lodders 2003). 
In contrast,
the Jones et al.\ (2013) model
consumed $\sidust\approx50\ppm$.
While the ZDA model used fewer
Si atoms ($\sidust\approx36\ppm$)
than the WD01 model
($\sidust\approx48\ppm$),
the former produces much elss extinction
at $\lambda > 1\mum$ than the latter
(see Figure~23.11 in Draine 2011).

Historically, ices were among the first grain species
suggested to be present in the ISM and responsible for
the observed extinction (see Li 2005b).
Oort \& van de Hulst (1946) proposed
the ``dirty ice'' model consisting of
saturated molecules
such as H$_2$O, CH$_4$, and NH$_3$.
In the 1960s, the first attempt to search for
the 3.1$\um$ feature of H$_2$O ice
in the diffuse ISM was unsuccessful
(Danielson et al.\ 1965, Knacke et al.\ 1969).
This led to the abandonment of
the ``dirty ice'' model and the proposition
of ``organic refractory'' resulting from
the photoprocessing of ``dirty ices''
as an interstellar grain component
(Greenberg et al.\ 1972, 1995).\footnote{%
  The core-mantle model of
  Li \& Greenberg (1997) assumes that
  the silicate core is coated by
  a mantle of organic refractory material.
  Whittet (2010b) suggested that the organic
  refractory matter may potentially
  account for some of the missing O atoms.
  }
Based on the then ``surplus'' of O, C, N elements,
Greenberg (1974) speculated that these ``excessive''
elements could be bound in ``interstellar snowballs''
ranging in size from baseballs to comets.
To the best of our knowledge,
the ZDA model is the only contemporary model
in which sub-$\mu$m-sized $\water$ ice was
included as a grain component and accounted for
$\simali$4.4\% of the total dust mass.

A crucial issue for the hypothesis of
$\mu$m-sized $\water$ ice grains as
the reservoir of the missing O atoms
is how they form and survive in the ISM.
In dense molecular clouds, $\water$ forms
through hydrogenation of O on the surfaces
of sub-$\mu$m-sized silicate and carbonaceous grains
and in this way, a thick ice mantle could be built up
(Boogert et al.\ 2015).\footnote{%
  One would expect an ice grain to
  have a sub-$\mu$m-sized nucleation
  core of silicate or graphite.
  But for the $\mu$m-sized ice grains
  of interest here, the effects of the core
  on the extinction and emission
  are negligible since the core only accounts for
  $\simali$$\left(0.1/4\right)^3\approx1.6\times10^{-5}$
  of the volume of a grain of $a=4\mum$.
  }
In contrast, ice mantles cannot be built up
in the diffuse ISM since they will be rapidly
removed by photosputtering
(Barlow 1978a,b; Draine \& Salpeter 1979).
We note that in the diffuse ISM,
not only $\water$ ice but also
silicate and graphite are destroyed
at a rate faster than their production
(McKee 1989). This led Draine (1990) to
conclude that the bulk of the solid material
in grains actually condensed
in the ISM rather than in stellar outflows.
Draine (1990) argued that there must be rapid exchange
of matter between the diffuse ISM and molecular clouds
since the bulk of grain growth can proceed rapidly
only in dense regions.
Depending on the posited mass exchange scenario,
this implies a turnover time
($\taumc$) of $\simali$$3\times10^6$--$2\times10^7\yr$
for molecular clouds (Draine 1990).
If the photosputtering lifetimes ($\taupd$) of
$\mu$m-sized $\water$ ice grains
are not shorter than $\taumc$,
the ice grain model might be viable:
before they are destroyed by photosputtering,
very large ice grains in the diffuse ISM
are continuously replenished
by the freshly-condensed ice grains
formed in dense molecular clouds
through accretion and coagulation.

Let $\Ypd(\lambda)$ be the photosputtering yield
of $\water$ ice
(i.e., the number of $\water$ molecules desorbed
per absorbed UV photon of wavelength $\lambda$).
We derive the photodesorption rate of H$_2$O ice
as following:
%
\begin{equation}\label{eq:dNdt}
\Ndot \equiv dN/dt = 
\int \Ypd(\lambda)
\frac{\Cabs(a,\lambda)\times4\pi J_\lambda}
     {hc/\lambda}\,
d\lambda ~~,
\end{equation}
where $h$ is the Planck constant,
$c$ is the speed of light,
$\Cabs(a,\lambda)$ is the absorption cross section
of $\water$ ice grain of radius $a$
at wavelength $\lambda$,
and $J_\lambda$ is the intensity of the MMP83 ISRF.
We calculate the photosputtering lifetimes
of H$_2$O ices of radii $a$ from
%
\begin{equation}\label{eq:taupd}
\taupd(a) = \frac{1}{\Ndot}
         \frac{4\pi a^3\,\rhoice}
              {3\muice\mH} ~~.
\end{equation}
Westley et al.\ (1995) experimentally measured
the photodesorption yield of H$_2$O ice
exposed to Lyman-$\alpha$ photons
(10.2\,eV or 1216$\Angstrom$)
at temperature $T$
to be $\Ypd \approx Y_0 + Y_1\exp\left(-E/kT\right)$,
where $k$ is the Boltzmann constant,
$Y_0=0.035\pm0.002$, $Y_1=0.13\pm0.10$,
and $E = (29\pm6)\times10^{-3}\eV$.
For $\water$ ice grains of radii $a=4\mum$
and temperature $T=8.9\K$
(see Figure~\ref{fig:Teq}),
we estimate the photosputtering lifetime
to be $\taupd\approx5.8\times10^6\yr$
in the diffuse ISM.
This lifetime may be underestimated
since the photodesorption yield of
Westley et al.\ (1995) could have been
overestimated as their data were obtained
through laser beam irradiation which might
have induced local point heating and thus
adding some sublimation to the pure sputtering
effect.
More recently, \"{O}berg et al.\ (2009)
experimentally determined
the photodesorption yields of H$_2$O ice
using a hydrogen discharge lamp ($\simali$7--10.5$\eV$).
They derived an average photodesorption yield of
$\Ypd\approx1.3\times10^{-3} + 3.2\times10^{-5}\times T$.
This gives a photosputtering lifetime
of $\taupd\approx7.4\times10^6\yr$
for $\water$ ice grains of $a=4\mum$ at $T=8.9\K$.
Moreover, Andersson et al.\ (2006) modeled
the photodesorption process at the atomic level
by simulating the interaction between photons
of $\simali$8--9.5$\eV$ and ice surfaces.
They estimated the photodesorption yield
to be $\Ypd\approx4\times10^{-4}$ for amorphous ice.
This leads to a photosputtering lifetime
of $\taupd\approx2.9\times10^7\yr$
for $\water$ ice grains of $a=4\mum$.
Therefore, it seems likely that the diffuse-phase
$\mu$m-sized $\water$ ice grains are continuously
replenished on a timely fashion
by the dense-phase materials
through the turnover of molecular clouds
within a timescale of
$\taumc$\,$\simali$$3\times10^6$--$2\times10^7\yr$
before they are destroyed by photosputtering
in the diffuse ISM.
In contrast, a sub-$\mu$m-sized ice grain
(coated on a silicate or graphite core)
responsible for the 3.1$\mum$ absorption feature
in dense clouds will be quickly removed by
photosputtering in diffuse clouds
at a rate faster than that of an $a=4\mum$ ice grain
by a factor of
$\simali$$\left(4/0.1\right)^3\approx6.4\times10^4$.

Finally, we note that the ``missing'' O atoms
do not {\it all} have to be tied up in $\water$ ice grains.
As mentioned earlier, some of the ``missing'' O atoms
could be hidden in organic refractories (Whittet 2010b).
On the other hand, the observed flat extinction
at $\simali$3--8$\mum$ could also {\it partly} be caused by
other dust components, e.g., $\mu$m-sized graphite
(see Wang, Li, \& Jiang 2015).

%
\section{Summary}\label{sec:summary}
While the ISM seems to be short of the element C
to form a sufficient amount of carbonaceous dust
(together with silicate dust)
to account for the observed extinction,
the element O seems overabundant and
as many as $\simali$160 O atoms
(per 10$^6$ H atoms) are unaccounted for
by their presence in gas and dust.
We have examined the hypothesis of
$\mu$m-sized $\water$ ice grains
as the reservoir of the missing O atoms.
It is found that the diffuse ISM has no difficulty in
accommodating $\simali$160$\ppm$ of O/H
in $\mu$m-sized $\water$ ice grains,
confirming the earlier suggestions made by
Jenkins (2009) and Poteet et al.\ (2015).
With a radius of $\simali$4$\mum$,
these grains are ``gray'' in the UV/optical
and contribute very little
to the UV/optical extinction.
The 3.1$\mum$ O--H stretching feature of
these grains is significantly suppressed,
consistent with the nondetection of this feature
in the diffuse ISM.
They absorb and scatter effectively
in the mid-IR and are capable of accounting
for the observed flat extinction
at $\simali$3--8$\mum$.
Being relatively transparent in the UV/optical,
they do not absorb much
and therefore they are cold
and mainly emit in the far-IR
at $\lambda\gtsim200\mum$,
accounting for only $\simali$0.34\%
of the total IR power of
the Galactic diffuse ISM.
With a photosputtering lifetime of
$\taupd$\,$\approx$\,$5.8\times10^6$--$2.9\times10^7\yr$
longer than or comparable to
the turnover timescale of molecular clouds
of $\taumc$\,$\approx$\,$3\times10^6$--$2\times10^7\yr$
implied by the observed large depletions
of Si and Fe elements in the diffuse ISM,
$\water$ ice grains of radii $a\gtsim4\mum$
could be present in the diffuse ISM
through rapid exchange of material
between dense molecular clouds
where they form
and diffuse clouds
where they are destroyed by
photosputtering.
%

\section*{Acknowledgements}
We thank A.C.A.~Boogert,
B.T.~Draine, G.M.~Mu\~noz Caro, 
K.I.~\"Oberg, A.N.~Witt, 
and the anonymous referee
for helpful comments/suggestions.
This work is
supported by NSFC 11173007, 11373015,
973 Program 2014CB845702,
NSF AST-1109039, and NNX13AE63G.
S.W. acknowledges support from
the China Scholarship Council (No.\ 201406040138).



\bsp
\label{lastpage}


\begin{thebibliography}{}
\bibitem[]{}Anders, E., \& Grevesse, N. 1989,
                 Geochim. Cosmochim. Acta, 53, 197
\bibitem[]{}Andersson, S., Al Halabi, A., Kroes, G. J.,
                 \& van Dishoeck, E. F. 2006, J. Chem. Phys., 124, 4715
\bibitem[]{}Arendt, R. G., et al., 1998, \apj, 508, 74
\bibitem[]{}Asplund, M., Grevesse, N., Sauval, A.J., \& Scott, P.\
                 2009, ARA\&A, 47, 481
\bibitem[]{}Barlow, M. J.\ 1978a, \mnras, 183, 397
\bibitem[]{}Barlow, M. J.\ 1978b, \mnras, 183, 417
\bibitem[]{}Bohren, C.F., \& Huffman, D.R.\ 1983, 
            Absorption and Scattering of Light by 
            Small Particles (New York: Wiley)
\bibitem[]{}Boogert A., Gerakines P., Whittet D., 2015, \araa, in press
\bibitem[]{}Cameron, A. G. W. 1973, Space Sci. Rev., 15, 121
\bibitem[]{}Cameron, A. G. W. 1982, in Essays in Nuclear Astrophysics,
                  ed. C. A. Barnes, D. D. Clayton, \& D. N. Schramm
                  (Cambridge: Cambridge Univ. Press), 23
\bibitem[]{}Cardelli, J. A., Clayton, G. C., \& Mathis, J. S. 1989,
                 \apj, 345, 245 (CCM)
\bibitem[]{}Cardelli, J. A., Meyer, D. M., et al., 1996, \apj, 467, 334
\bibitem[]{}Chiappini, C., Romano, D., \& Matteucci, F.\ 
            2003, MNRAS, 339, 63 
\bibitem[]{}Danielson, R. E., Woolf, N. J., \& Gaustad, J. E.\ 1965,
                 \apj, 141, 116
\bibitem[]{}Draine, B. T. 1989,
                 in Infrared Spectroscopy in Astronomy,
                 ed. B. H. Kaldeich (Paris: ESA Publ. Division), 93
\bibitem[]{}Draine, B. T. 1990, in ASP Conf. Ser.12,
                 The Evolution of the Interstellar Medium,
                 ed. L. Blitz (San Francisco, CA: ASP), 193
\bibitem[]{}Draine, B. T. 2011,
                 Physics of the Interstellar and Intergalactic Medium
                 (Princeton, NJ: Princeton Univ. Press)
\bibitem[]{}Draine, B. T., \& Lee, H. M. 1984, \apj, 285, 89
\bibitem[]{}Draine, B. T., \& Li, A. 2001, \apj, 551, 807
\bibitem[]{}Draine, B. T., \& Li, A. 2007, \apj, 657, 810
\bibitem[]{}Draine, B. T., \& Salpeter, E. E.\ 1979, \apj, 231, 438
\bibitem[]{}Flaherty, K. M., Pipher, J. L., Megeath, S. T.,
                 et al., 2007, \apj, 663, 1069
\bibitem[]{}Field, G. B. 1974, \apj, 187, 453
\bibitem[]{}Finkbeiner, D. P., Davis, M., \& Schlegel, D. J. 1999,
                 \apj, 524, 867
\bibitem[]{}Gao, J., Jiang, B.W., \& Li, A., 2009, \apj, 707, 89
\bibitem[]{}Greenberg, J.~M., Yencha, A.~J., Corbett, J.~W.,
                 \& Frisch, H.~L.\ 1972, in Les spectres des astres
                dans l'infrarouge et les microondes
                (Li\'ege: Soci\'et\'e Royale de Sciences de Li\'ege), 425
\bibitem[]{}Greenberg, J. M.\ 1974, \apjl, 189, L81
\bibitem[]{}Greenberg, J. M., Li, A., Mendoza-Gomez, C. X., et al.\
                 1995, \apjl, 455, L177
\bibitem[]{}Indebetouw, R., et al. 2005, \apj, 619, 931
\bibitem[]{}Jenkins, E. B. 2009, \apj, 700, 1299
\bibitem[]{}Jiang, B.W., Gao, J., Omont, A., Schuller, F.,
                 \& Simon, G. 2006, \aap, 446, 551
\bibitem[]{}Jones, A. P., Fanciullo, L., K{\"o}hler, M., et al.\ 2013,
                  \aap, 558, A62
\bibitem[]{}Knacke, R. F., Cudaback, D. D., \& Gaustad, J. E.\ 1969,
                 \apj, 158, 151
\bibitem[]{}Li, A. 2004, in ASP Conf. Ser., 309, Astrophysics of Dust,
                  ed. Witt, A.N., Clayton, G.C., \& Draine, B.T.,
                  (San Francisco: ASP), 417
\bibitem[]{}Li, A. 2005a, \apj, 622, 965
\bibitem[]{}Li, A. 2005b, J. Phys.: Conf. Ser., 6, 229
\bibitem[]{}Li, A., \& Draine, B. T. 2001, \apj, 554, 778
\bibitem[]{}Li, A., \& Greenberg, J. M. 1997, \aap, 323, 566
\bibitem[]{}Lodders, K. 2003, \apj, 591, 1220
\bibitem[]{}Lutz, D. 1999,
                 in The Universe as Seen by ISO,
                 ed. P. Cox \& M. Kessler
                 (ESA Special Publ., Vol.~427; Noordwijk: ESA), 623
\bibitem[]{}Mathis, J. S. 1996, \apj, 472, 643
\bibitem[]{}Mathis, J. S., Mezger, P. G., \& Panagia, N. 1983,
                 \aap, 128, 212 (MMP83)
\bibitem[]{}Mathis, J. S., Rumpl, W., \& Nordsieck, K. H. 1977,
                 \apj, 217, 425
\bibitem[]{}McFadzean, A.D., Whittet, D.C.B., Longmore, A.J., 
                 et al.\ 1989, \mnras, 241, 873
\bibitem[]{}McKee, C. F. 1989, in IAU Symp. 135, Interstellar Dust,
                 ed. L. J. Allamandola \& A. G. G. M. Tielens
                 (Dordrecht : Reidel), 431
\bibitem[]{}Morton, D. C., Drake, J. F., Jenkins, E. B., et al.\ 1973,
                 \apjl, 181, L103
\bibitem[]{}Nieva, M. F., \& Przybilla. N. 2012, \aap, 539, 143
\bibitem[]{}Nishiyama, S., Tamura, M., Hatano, H., et al.\ 2009,
                 \apj, 696, 1407
\bibitem[]{} {\"O}berg, K. I., Linnartz, H., Visser, R.,
                 \& van Dishoeck, E. F.\ 2009, \apj, 693, 1209
\bibitem[]{}Oort, J. H., \& van de Hulst, H. C. 1946,
                 Bull. Astron. Inst. Netherlands, 10, 187
\bibitem[]{}Pendleton, Y.~J., Sandford, S.~A., Allamandola, L.~J., 
                 et al.\ 1994, \apj, 437, 683
\bibitem[]{}Planck Collaboration XVII 2014, \aap, 566, A55
\bibitem[]{}Poteet, C. A., Whittet, D. C. B., \& Draine, B. T. 2015,
                 \apj, 801, 110
\bibitem[]{}Press, W. H., Teukolsky, S. A., Vetterling, W. T., 
                 \& Flannery, B. P. 1992, Numerical Recipes in FORTRAN: 
                 The Art of Scientic Computing 
                 (2d ed.; Cambridge: Cambridge Univ. Press)
\bibitem[]{}Purcell, E. M. 1969, \apj, 158, 433
\bibitem[]{}Sandford, S.A., Pendleton, Y.J., \& Allamandola, L.J.\ 1995,
                 \apj, 440, 697
\bibitem[]{}Snow, T. P., \& Witt, A. N. 1995, Science, 270, 1455
\bibitem[]{}Snow, T. P., \& Witt, A. N.\ 1996, \apjl, 468, L65
\bibitem[]{}Sofia, U. J., Parvathi, V. S., et al., 2011, \apj, 141, 22
\bibitem[]{}Tielens, A. G. G. M., Wooden, D. H., Allamandola, L. J.,
                 et al.\ 1996, \apj, 461, 210
\bibitem[]{}Wang, S., Gao, J., Jiang, B. W., Li, A., \& Chen, Y. 2013,
                 \apj, 773, 30
\bibitem[]{}Wang, S., Li, A., \& Jiang, B. W. 2014,
                 Planet. Space Sci., 100, 32
\bibitem[]{}Wang, S., Li, A., \& Jiang, B. W. 2015, \apj, in press
\bibitem[]{}Warren, S. G.\ 1984, Appl. Opt., 23, 1206
\bibitem[]{}Weingartner, J. C., \& Draine, B. T. 2001, \apj,
                 548, 296 (WD01)
\bibitem[]{}Westley, M. S., Baragiola, R. A., Johnson, R. E.,
                 \& Baratta, G. A.\ 1995, Nature, 373, 405
\bibitem[]{}Whittet, D. C. B., Boogert, A. C. A., Gerakines, P. A., et al.
                 1997,  \apj, 490, 729
\bibitem[]{}Whittet, D. C. B., Pendleton, Y. J., Gibb, E. L., et al.\
                 2001, \apj, 550, 793
\bibitem[]{}Whittet, D. C. B.\ 2010a, LPI Contributions, 1538, 5194
\bibitem[]{}Whittet, D. C. B.\ 2010b, \apj, 710, 1009
\bibitem[]{}Zubko, V., Dwek, E., \& Arendt, R. G.\ 2004, \apjs, 152, 211


%
\end{thebibliography}
\end{document}